\documentstyle[preprint,aps,floats,epsfig]{revtex}

\newcommand{\be}{\begin{equation}}
\newcommand{\ee}{\end{equation}}
\newcommand{\bea}{\begin{eqnarray}}
\newcommand{\eea}{\end{eqnarray}}
\newcommand{\bml}{\begin{mathletters}}
\newcommand{\eml}{\end{mathletters}}

\begin{document}

\tighten

\preprint{DCPT-02/19}
\draft




\title{Remarks on the interaction between Born-Infeld solitons}
\renewcommand{\thefootnote}{\fnsymbol{footnote}}
\author{ Yves Brihaye\footnote{Yves.Brihaye@umh.ac.be}}
\address{Facult\'e des Sciences, Universit\'e de Mons-Hainaut,
 B-7000 Mons, Belgium}
\author{Betti Hartmann\footnote{Betti.Hartmann@durham.ac.uk}}
\address{Department of Mathematical Sciences, University
of Durham, Durham DH1 3LE, U.K.}
\date{\today}
\setlength{\footnotesep}{0.5\footnotesep}

\maketitle
\begin{abstract}
We consider the Abelian Higgs model as well as the
SU(2) Georgi-Glashow model in
which the gauge field action is replaced by a non linear Born-Infeld
action. We study soliton solutions arising in these models, namely the
vortex and monopole solutions, respectively. We construct 
formulas which provide good approximations 
for the mass of the Born-Infeld deformed solitons
using only the data of the undeformed solutions. 
The results obtained indicate that in the self-dual
limit, the Born-Infeld
interaction leads to bound vortices, 
while for monopoles it gives rise to repulsion.
\end{abstract}


\renewcommand{\thefootnote}{\arabic{footnote}}

\section{Introduction}
In recent years it became apparent that when studying 
low energy effective actions of string theory, the part of
the Lagrangian containing 
the abelian Maxwell field strength tensor
and its non-abelian counterpart in  Yang-Mills
field theories, respectively, must be replaced by a corresponding
(resp. abelian and non-abelian) Born-Infeld term. This idea is not new
and was first suggested by Born and Infeld in the 1930s \cite{bi}
to get rid of singularities associated with  point-like charges
in electrodynamics. A subtle point in the generalisation
to non-abelian Born-Infeld (NBI) theory is how to
specify the trace over the gauge group generators.
From the viewpoint of string theory
the {\it symmetrized} trace \cite{tse} seems more favourable than
the {\it ordinary} trace. In general, the Lagrangian
involving a {\it symmetrized} trace 
is only known in a perturbative expansion \cite{moreno}. 
Recently, however, an explicit expression of the SU(2)
NBI action for static, spherically symmetric, magnetic
configurations involving a {\it symmetrized} trace was constructed \cite{galtsov1}.
It was found that the qualitative results are in good agreement with
the ones obtained previously \cite{galtsov2} using the {\it ordinary} trace.\  

A number of classical field theory model 
have been studied with respect
to the effects of the Born-Infeld (BI) interaction. It was
found that in SU(2) Yang-Mills theory the presence of the
Born-Infeld term leads to the existence of
particle-like solutions, so called ''glueballs'' \cite{galtsov2},
which are absent in the standard case (i.e. without BI interaction). 
The reason for this
is that similar to gravity in the Einstein-Yang-Mills model \cite{bartnik}, 
the Born-Infeld term breaks the scale invariance and thus admits
finite energy solutions.
For models in which soliton-type solutions already exist 
in the standard case, it is of interest to study the BI deformation
of these configurations. This has been done for a number of models
including the Abelian Higgs model \cite{bm} and the SU(2)
Georgi-Glashow model \cite{gpss}. 

Both, the  Abelian Higgs model \cite{no}
and the Georgi-Glashow model \cite{thooft} admit soliton
solutions characterized by 
an integer of topological origin and possess a so-called
''self-dual'' (BPS) limit for a specific value of the
Higgs self-coupling constant. In this limit, the second order
Euler-Lagrange equations can be replaced by a set of first order (''self-dual'')
equations and the mass of the $n$-soliton solution is given by
$M(n) = n M(n=1)$ indicating that no
interaction between the solitons exists \cite{rebbi}. The 
repulsion due to the long range gauge field is exactly
compensated by the attraction of the Higgs field which, because of 
its masslessness, is also long range.

A natural question arising in the study of the Born-Infeld interaction
is whether it can lead to attraction between 
vortices and monopoles, respectively.  In the case of vortices, 
it was found that in the self-dual limit
the Born-Infeld interaction leads to attraction \cite{bm}.
For monopoles the question has not be answered yet and one of the aims of this
paper is to give an indication of whether the Born-Infeld interaction
similar to gravity \cite{hkk} can lead to attraction.\

Here we present a formula which provides a good approximation
for the energy of  the BI-deformed solitons using only the data
of the standard case. In Section II, we study this approximation
for the Abelian Higgs model, in section III for the Georgi-Glashow
model. We give our conclusion in section IV.

\section{The Abelian Higgs model}
The Lagrangian of the Abelian Higgs model reads \cite{no}:
\begin{equation}
{\cal L}=-\frac{1}{4}F_{\mu\nu}F^{\mu\nu}-
\frac{1}{2}D_{\mu}\Phi D^{\mu}\Phi^{*}-\frac{\lambda}{4}(\Phi^{*}\Phi-v^2)^2
\end{equation}
where $D_{\mu}=\partial_{\mu}-ieA_{\mu}$, $A_{\mu}$ is the U(1) gauge potential
with coupling $e$ and $\Phi$ is a complex Higgs field with vacuum expectation value $v$ and
self-coupling constant $\lambda$.

This model has soliton solutions, namely 
the Nielsen-Olesen vortices \cite{no} and
$n$-soliton solutions can be constructed
with an appropriate axially symmetric ansatz~:
\begin{equation}
\Phi=vf(\rho)e^{i\varphi}  \ ,  \ A_{\mu}dx^{\mu}=\frac{1}{e}(n-P(\rho))d\varphi
\end{equation}
The self-dual case
is given for $\alpha=\frac{e^2}{\lambda}=2$.\
In the Born-Infeld version of this theory \cite{bm},
the part of the Lagrangian
containing the Maxwell field strength tensor $F_{\mu\nu}$
is replaced by the corresponding 
Born-Infeld (BI) expression with BI coupling $\beta$~:

\begin{equation}
-\frac{1}{4}F_{\mu\nu} F^{\mu\nu} \longrightarrow 
\beta^2\left( 1-\sqrt{1+\frac{1}{2\beta^2}F_{\mu\nu} F^{\mu\nu}-
\frac{1}{16\beta^4}(F_{\mu\nu} \tilde{F}^{\mu\nu})^2}\right)
\label{ft}
\end{equation} 
For $\beta^2 >> 1$, the square root can be evaluated as follows:
$$
 \sqrt{1+\frac{1}{2\beta^2}F_{\mu\nu} F^{\mu\nu}-\frac{1}{16\beta^4}
(F_{\mu\nu} \tilde{F}^{\mu\nu})^2}\approx   \\ \nonumber 
$$
\begin{equation} 
  1+
\frac{1}{4\beta^2}F_{\mu\nu} F^{\mu\nu}
-\frac{1}{32\beta^4}
[(F_{\mu\nu} F^{\mu\nu})^2+(F_{\mu\nu} 
\tilde{F}^{\mu\nu})^2]\pm O(\beta^{-6}) 
\label{fapprox}
\end{equation} 
For  $\beta^2\rightarrow\infty$, the standard expression
of the U(1) theory on the lhs of (\ref{ft})
is recovered. In the limit of 
static and purely magnetic solutions, the part containing
the dual field strength tensor $\tilde{F}_{\mu\nu}$ vanishes.

In this paper, we are interested in static, finite energy solutions of the
full BI equations
with classical mass denoted by $M_{bi}(\beta^2)$.
Using (\ref{fapprox}), we obtain the following approximation 
for this mass~:
\be
\label{approx}
M_{bi}(\beta^2) \approx  M_0 - \frac{1}{32\beta^2} \int d^3 x 
(F_{(0)j k} F_{(0)}^{j k})^2 
\equiv M_0 - \frac{1}{\beta^2} \Delta_{bi}
\ee
Both, the mass of the standard soliton $M_0$ and the field strength
tensors $F_{(0) j k}$ are computed from  the standard
equations. In other words, we approximate the mass of the BI-soliton by
the $\beta^{-2}$- correction of the BI Lagrangian density using only the
solution data of the standard U(1) theory.\

For the self-dual case,
the first order equations read:
\begin{equation}
f^{'}=\frac{Pf}{x} \ , \ \ P^{'}=x^2(f^2-1)
\end{equation}
In this limit, the mass  
$M_0$ of the Nielsen-Olesen vortex and the correction $\Delta_{bi}$
are given as follows~:
\be
     M_0 = \int dx x [ \frac{(P')^2}{x^2}
                       + (f')^2 + \frac{P^2 f^2}{x^2}
                       + \frac{1}{4} (1-f^2)^2  ]
\label{mass}
\ee 
and
\be
      \Delta_{bi} = \frac{1}{2} \int dx x [\frac{P'}{x}]^4
\label{correction}
\ee
where $x$ is a dimensionless coordinate $x=ev\rho$ and the prime denotes
the derivative with respect to $x$. With the 
choice of coordinates, the {\it mass per winding number} of the $n$-soliton 
solution in the self-dual 
case is given by $\frac{M_0}{n}=1$ (in units of $2\pi \frac{v^2}{\alpha}$).\

Evaluating the correction integral numerically, we obtain
$\Delta_{bi}(n=1) \approx 0.0118$ and $\Delta_{bi}(n=2)/n \approx 0.02037$.
In the table below, we compare the {\it masses per winding number} 
obtained with this correction
and the {\it masses per winding number}
$M_{bi}(\beta^2)(n)/n$ computed in \cite{bm} with the full equations.
\begin{center}
\begin{tabular}{|l|l|l|l|l|}
\hline
$\beta^2$ & $M_{bi}(\beta^2)(n=1)$ & $M_0-\frac{\Delta_{bi}}{\beta^2}(n=1)$
& $M_{bi}(\beta^2)(n=2)/2$ & $(M_0-\frac{\Delta_{bi}}{\beta^2}(n=2))/2$ \\
\hline 
$ \infty$    & $1.0 $  & $1.0 $ & $1.0 $ & $1.0$\\
$100.$         &$0.99988 $    & $0.99988 $ &$0.99979 $ &$0.99979$\\
$10.$         &$0.99880 $    & $0.99882 $ &$0.99790 $ &$0.99796$\\
$5.$          &$0.99760 $    & $0.99764 $ &$0.99585 $ &$0.99592$ \\
$1.$          &$0.9870 $    & $0.9882 $ &$0.97780 $ &$0.97963 $\\
$0.5$         &$0.97038 $    & $0.9764$ &$0.94980 $ & $0.95926 $\\
\hline
\end{tabular}
\end{center}
Clearly, the approximation is very good for large values of $\beta^2$ and
as expected gets worse for decreasing $\beta^2$, since the approximation
used in (\ref{fapprox}) is not valid anylonger and higher order terms have
to be taken into account.\

The quality of this approximation for large $\beta^2$ also provides
a good estimation for the difference between the
{\it energy per winding number} of the $n=1$ and the $n=2$ vortex~:
\be
   \delta M  = (M_0-\frac{\Delta_{bi}}{\beta^2}(n=1))-
(M_0-\frac{\Delta_{bi}}{\beta^2}(n=2))/2 \approx 0.00875 \frac{1}{\beta^2} 
\ee
This clearly confirms the results in \cite{bm}, especially that the Born-Infeld
interaction stabilizes the axially symmetric 2-vortex 
solution for $\alpha=2$.

\section{The Georgi-Glashow model}
The Lagrangian of the SU(2) Yang-Mills-Higgs theory with the
Higgs field in the adjoint representation (the so-called 
Georgi-Glashow model) reads:
\begin{equation}
{\cal L}=-\frac{1}{4}F_{\mu\nu}^{a}F^{\mu\nu,a}-
\frac{1}{2}D_{\mu}\Phi^{a} D^{\mu}\Phi^{a}
-\frac{\lambda}{4}(\Phi^{a}\Phi^{a}-v^2)^2 \ , \ a=1, 2, 3
\end{equation}
where the field strength tensor $F_{\mu\nu}^{a}$ and the covariant 
derivative of the Higgs field $D_{\mu}\Phi^{a}$ are given as follows:
\begin{equation}
F_{\mu\nu}^{a}=\partial_{\mu}A_{\nu}^{a}-\partial_{\nu}A_{\mu}^{a}+
e\varepsilon_{abc}A_{\mu}^{b}A_{\nu}^{c} \ , \ \ \
D_{\mu}\Phi^{a}=\partial_{\mu}\Phi^{a}+
e\varepsilon_{abc}A_{\mu}^{b}\Phi^{c}
\end{equation}
The soliton solutions of this model
are magnetic monopoles \cite{thooft}.\

For a similar approximation as in (\ref{approx}), we 
expand the {\it ordinary} trace version of the non-abelian Born-Infeld (NBI)
model. This is a straightfoward generalisation of (\ref{fapprox})
replacing $F_{\mu\nu}F^{\mu\nu}$ by $F^a_{\mu\nu}F^{a\mu\nu}$. We also give
the first term in the approximation of the {\it symmetrized} trace version
extending the results of \cite{galtsov1}.
\subsection{Spherically symmetric monopoles}
For the gauge and Higgs fields, we use the purely magnetic hedgehog ansatz
($a=1, 2, 3$) \cite{thooft}~:
\begin{equation}
{A_r}^a={A_t}^a=0 \ , \  \  \ {A_{\theta}}^a= \frac{1-K(r)}{e} {e_{\varphi}}^a
\ , \ \ \ \ 
{A_{\varphi}}^a=- \frac{1-K(r)}{e}\sin\theta {e_{\theta}}^a
\ , 
\label{hpmono1}
\end{equation}
\begin{equation}
{\Phi}^a=v H(r) {e_r}^a
\label{hpmono2}
\ . \end{equation}

The spherically symmetric $n=1$ Born-Infeld monopoles were constructed in \cite{gpss}
both by using the {\it ordinary} and the {\it symmetrized} trace.
It was found that for large values of the Born-Infeld coupling $\beta^2$,
the profiles of the functions don't differ significantly from those of the
't Hooft-Polyakov monopole. Equally, the mass only depends
slightly on $\beta^2$. It was found, however, that 
the falloff of the Higgs field in the limit of vanishing Higgs coupling
now is given by $c(\beta^2)/r$, where $c$ is a constant depending
on $\beta^2$ with $c=1$ in the limit $\beta^2\rightarrow\infty$.
For the {\it ordinary} trace model,
a critical value of $\beta^2$ was found, $\beta^2_{cr}$, 
such that for $\beta^2 < \beta^2_{cr}$ no solutions exist. For vanishing Higgs 
self-coupling this was determined to be $\beta^2_{cr}=0.168$. \

In the case of spherically symmetric solutions in the BPS limit, the
first order (self-dual) differential equations read:
\begin{equation}
K^{'}=HK \ , \ \ x^2H^{'}=(K^2-1)
\end{equation}
In the BPS limit, the non-abelian counterparts
to the integrals (\ref{mass}) and (\ref{correction}) read~:
\be
  M_0(n=1) = \int dx x^2 [ \frac{(K')^2}{x^2} + \frac{(K^2-1)^2}{2 x^4}
                       + \frac{1}{2}(H')^2 + \frac{K^2 H^2}{x^2} ]
\ee
  \be
      \Delta_{bi}^{tr} = \frac{1}{2} 
\int dx x^2 [\frac{(K')^2}{x^2}  + \frac{(K^2-1)^2}{2 x^4}]^2
\label{ot}
\ee
for the {\it ordinary} trace denoted by $tr$, and
\begin{equation}
\Delta_{bi}^{Str} = \frac{1}{2} 
\int dx x^2 [\frac{2}{3}(\frac{(K')^2}{x^2})^2  + (\frac{(K^2-1)^2}{2 x^4})^2
+ \frac{2}{3}\frac{(K^2-1)^2}{2 x^4}\frac{(K')^2}{x^2}]
\label{st}
\end{equation}
for the {\it symmetrized} trace denoted by $Str$.
The mass $M_0$ is given in unit of $4\pi\frac{v}{e}$, $x=evr$ and the prime
denotes the derivative with respect to $x$.\
The numerical evaluation of the integrals (\ref{ot}) and (\ref{st})
for the $n=1$-monopole
gives $\Delta_{bi}^{tr} \approx 0.00919$ and $\Delta_{bi}^{Str} \approx 0.00515$.
Similar as in  the case of vortices, this provides a
very good approximation of the BI-monopole mass for large $\beta^2$
as the table below demonstrates~:
\begin{center}
\begin{tabular}{|l|r|r|r|r|}
\hline
$\beta^2$ & $M_{bi}^{tr}(\beta^2)$ & $M_0-\Delta_{bi}^{tr}/\beta^2$  &
$M_{bi}^{Str}(\beta^2)$ & $M_0-\Delta_{bi}^{Str}/\beta^2$   \\
\hline 
$ \infty$     &$1.0$         & $1.0$  & $1.0$ & $1.0$ \\
$100.$         &$0.99991$    & $0.99991$ & $0.99994$ & $0.99994$  \\
$10.$         &$0.99907$    & $0.99908$ & $0.99948$  & $0.99948$   \\
$5.$          &$0.99813$    & $0.99816$ & $0.99895$ & $0.99897$\\
$1.$          &$0.99012$    & $0.99080$ & $0.99430$  & $0.99485$  \\
$0.5$         &$0.97829$    & $0.98162$ & $-$  & $-$\\
\hline
\end{tabular}
\end{center}
$M_{bi}^{tr}(\beta^2)$ was computed using the full Born-Infeld equations.
The integration of the full equations in the {\it symmetrized} trace
version becomes very involved and thus the mass $M_{bi}^{Str}(\beta^2)$
denotes the mass of the BI monopole computed from the 
Lagrangian involving terms up to order $\beta^{-2}$. The reason why we don't
give the numbers for $\beta^2=0.5$ in the {\it symmetrized} trace
version is that we couldn't integrate the equations for $\beta^2 < 0.6$. This leaves
us with the assumption that for the equations derived
from the {\it symmetrized} trace Lagrangian involving terms
up to order $\beta^{-2}$ a $\beta^2_{cr}$ also exists and that in the case studied
here $\beta^2_{cr}\approx 0.6$. However, further
investigation of this is out of the
aim of this paper.  \

\subsection{Axially symmetric monopoles}
The axially symmetric Ansatz for the gauge fields is given by \cite{rr}~: 
\begin{eqnarray}
A_\mu dx^\mu =\frac{1}{2} A_\mu^a \tau^a dx^\mu &=& 
\frac{1}{2er} [ \tau^n_\varphi 
 ( H_1 dr + (1-H_2) r d\theta ) \nonumber  \\
 &-& n ( \tau^n_r H_3 + \tau^n_\theta (1-H_4))
  r \sin \theta d\varphi ]
\ , \label{axg}
\end{eqnarray}
while for the Higgs field it reads
\begin{equation}
\Phi= \Phi^a\tau^a = \left(\Phi_1 \tau_r^{n}+\Phi_2 \tau_\theta^{n}\right)
\label{axh}
\  \end{equation}
where the matter field functions $H_1$, $H_2$, $H_3$, $H_4$, $\Phi_1$ and
$\Phi_2$ depend only on $r$ and $\theta$.
The symbols $\tau^n_r$, $\tau^n_\theta$ and $\tau^n_\varphi$
denote the dot products of the cartesian vector
of Pauli matrices, $\vec \tau = ( \tau^1, \tau^2, \tau^3) $,
with the spatial unit vectors
\begin{eqnarray}
\vec e_r^{\, n}      &=& 
(\sin \theta \cos n \varphi, \sin \theta \sin n \varphi, \cos \theta)
\ , \nonumber \\
\vec e_\theta^{\, n} &=& 
(\cos \theta \cos n \varphi, \cos \theta \sin n \varphi,-\sin \theta)
\ , \nonumber \\
\vec e_\varphi^{\, n}   &=& (-\sin n \varphi, \cos n \varphi,0) 
\ , \label{rtp} \end{eqnarray}
For $H_1=H_3=\Phi_2=0$, $H_2=H_4=K(r)$, $\Phi_1=H(r)$ and $n=1$, the Ansatz
(\ref{hpmono1}), (\ref{hpmono2})
for the 't Hooft-Polyakov monopole \cite{thooft} is recovered. The self-dual
equations read:
\begin{equation}
F_{ij}^{a}=\pm \varepsilon_{ijk}\sqrt{-g} D^{k,a}\Phi=
\pm\varepsilon_{ijk} r^2 \sin\theta D_{k}^{a}\Phi g^{kk}
\end{equation} 

It would be interesting to construct the Born-Infeld
$n=2$-monopole and compare its mass with the one of the $n=1$ BI-monopole
\cite{gpss}
as well as with the standard $n=2$-monopole \cite{rr,kkt}.
This can be done only by using the full axially symmetric ansatz (\ref{axg}),
(\ref{axh}).  
The occurence of non-polynomial
terms in  the Born-Infeld action leads to the fact that
the corresponding partial differential equations are not manifestly
elliptic (-mixed derivatives of the fields are unavoidable-).
The numerical integration seems to be very 
involved and is left for future work. 
However, encouraged by the results described above, 
we propose an estimation for the
energy of the $n=2$-axially symmetric Born-Infeld monopole.
For the $n=2$ monopole, the relevant integrals in the BPS limit read:
\begin{eqnarray}
M_0(n) &= & 
\frac{1}{2}\int \int \sin\theta d\theta dx x^2 \left[
                            \frac{1}{x^2}\left( 
(x\partial_x \Phi_1 +H_1 \Phi_2)^2
                             +(x\partial_x \Phi_2 -H_1 \Phi_1)^2
                             +(\partial_\theta \Phi_1 -H_2 \Phi_2)^2
                             \right.  \right.
\nonumber \\
& & 
        \left.    \left.    +(\partial_\theta \Phi_2 +H_2 \Phi_1)^2
            + n^2 (H_4 \Phi_1 +(H_3+{\rm cot}\theta )\Phi_2)^2 \right)
                +\frac{1}{x^4} \left\{
  \left(x \partial_x H_2 + \partial_\theta H_1\right)^2 
 \right. \right.
\nonumber \\
& &  
+  \left.  \left.
   n^2 \left(
  \left(  x \partial_x H_3 - H_1 H_4 \right)^2
+ \left(x \partial_x H_4 + H_1 \left( H_3 + {\rm cot} \theta \right)
    \right)^2 \right. \right. \right.
\nonumber \\
      &  & \left. \left.  \left.
 + \left(\partial_\theta H_3 - 1 + {\rm cot} \theta H_3 + H_2 H_4
     \right)^2 +
  \left(\partial_\theta H_4 + {\rm cot} \theta \left( H_4-H_2 \right) 
   - H_2 H_3 \right)^2 \right) \right\}  \right]
 \label{edens} \end{eqnarray} 
and
\begin{eqnarray}
\Delta_{bi}^{tr}&=&\frac{1}{8} \int \int \sin\theta d\theta dx
\frac{1}{x^6} \left\{
  \left(x \partial_x H_2 + \partial_\theta H_1\right)^2 +
   n^2 \left[
  \left(  x \partial_x H_3 - H_1 H_4 \right)^2
 \right. \right.
\nonumber \\
& &  
+ \left. \left.
    \left(x \partial_x H_4 + H_1 \left( H_3 + {\rm cot} \theta \right)
    \right)^2 + \left(\partial_\theta H_3 - 1 + {\rm cot} \theta H_3 + H_2 H_4
     \right)^2  \right. \right.
\nonumber \\
      &  & \left. \left.
  +
  \left(\partial_\theta H_4 + {\rm cot} \theta \left( H_4-H_2 \right) 
   - H_2 H_3 \right)^2 \right] \right\}^{2}
\label{axtr}
\end{eqnarray}
We derive the expression for $\Delta^{Str}_{bi}$ in the case
of axially symmetric monopoles in the Appendix. We obtain:
\begin{eqnarray}
\Delta_{bi}^{Str}&=&\frac{1}{16} \int \int \sin\theta d\theta dx
\frac{1}{x^6} \left\{
  \left(x \partial_x H_2 + \partial_\theta H_1\right)^4 +
   n^4 \left[
  \left(  x \partial_x H_3 - H_1 H_4 \right)^2
 \right. \right.
\nonumber \\
& &  
+ \left. \left.
     \left(\partial_\theta H_3 - 1 + {\rm cot} \theta H_3 + H_2 H_4
     \right)^2  \right]^2 +
 n^4 \left[
  \left(  x \partial_x H_4 - H_1 H_3+ {\rm\cot}\theta H_1 \right)^2
 \right. \right.
\nonumber \\
& &  
+ \left. \left.
     \left(\partial_\theta H_4 - H_2 H_3 - {\rm cot} \theta ( H_2- H_4)
     \right)^2  \right]^2 +
 \frac{4 n^4}{3} \left[
  \left(  x \partial_x H_3 - H_1 H_4 \right)\left(x \partial_x H_4+
H_1 H_3 + {\rm\cot}\theta H_1 \right) 
 \right. \right.
\nonumber \\
& &  
+ \left. \left.
     \left(\partial_\theta H_3 - 1 + H_2 H_4 - {\rm cot} \theta H_3\right)
\left(\partial_\theta H_4 - H_2 H_3 - {\rm cot} \theta (H_2 -H_4)
     \right)  \right]^2 + \frac{2 n^2}{3} \left(\partial_\theta H_1 + x \partial_x H_2\right)^2\cdot
 \right.
\nonumber \\
      &  &  \left.  
\left[ \left(x \partial_x H_3-H_1 H_4\right)^2
 + \left( \partial_\theta H_3- 1 + H_4 H_4 +{\rm cot} \theta H_3 \right)^2+
     \left( \partial_\theta H_4-H_2 H_3-{\rm cot} \theta (H_2-H_4)\right)^2 
\right. \right.
\nonumber \\
   & &
+   \left. \left.
 \left(x \partial_x H_4+H_1 H_3 +{\rm cot} \theta H_1\right)^2                            
   \right]
+ \frac{2 n^4}{3}\left[\left(x \partial_x H_3 - H_1 H_4\right)^2 +
\left( \partial_\theta H_3 - 1+  H_2 H_4 +{\rm cot} \theta H_3 \right)^2 \right]\cdot
\right. 
\nonumber \\
   & &
   \left. 
\left[\left(x \partial_x H_4 + H_1 H_3 + {\rm cot} \theta H_1\right)^2 +
\left( \partial_\theta H_4 -  H_2 H_3 -{\rm cot} \theta (H_2-H_4) \right)^2 \right]
\right\}
\label{axstr}
\end{eqnarray}

We find  $(1/n) \Delta_{bi}^{tr}=(1/2) \Delta_{bi}^{tr} \approx 0.00303$ and
$(1/n) \Delta_{bi}^{Str}=(1/2) \Delta_{bi}^{Str} \approx 0.00187$.
With this approximation, the  difference between the {\it mass per winding number} 
of the $n=1$ and the $n=2$ monopole in the BPS limit is given by~:
\be
 \delta M ^{tr}  = (M_0-\frac{\Delta_{bi}^{tr}}{\beta^2}(n=1))-
(M_0-\frac{\Delta_{bi}^{tr}}{\beta^2}(n=2))/2
\approx -0.006 \frac{1}{\beta^2} 
\ee
and
\be
 \delta M ^{Str}  = (M_0-\frac{\Delta_{bi}^{Str}}{\beta^2}(n=1))-
(M_0-\frac{\Delta_{bi}^{Str}}{\beta^2}(n=2))/2
\approx -0.003 \frac{1}{\beta^2} 
\ee
Note that with our choice of coordinates, the {\it mass per winding number} of the BPS monopoles
is given by $\frac{M_0}{n}=1$.\
 
If we rely on  the approximation
(\ref{approx}) and extend it to the case of multimonopoles, 
the above result indicates that the Born-Infeld interaction  
leads to repulsion of like-charged monopoles. To have further indication, we computed
the integrals (\ref{axtr}) and (\ref{axstr}) with the axially symmetric Ansatz \cite{kkt} for 
winding number up to $n=7$. The values are given below:

\begin{center}
\begin{tabular}{|l||l|l|l|l|l|l|l|}
\hline
$n$ & $1$ &$2$& $3$ &$4$& $5$& $6$& $7$\\
\hline
$ 10^{3}\cdot\Delta_{bi}^{tr}/n$ & $9.198 $ &$3.035$& $1.600$& $1.008$& $0.699$& $0.517$& $0.397$\\
\hline
$ 10^{3}\cdot\Delta_{bi}^{Str}/n$ & $5.156 $ &$1.869$& $1.064$& $0.705$& $0.506$& $0.385$& $0.303$\\
\hline
\end{tabular}
\end{center}
Clearly, with increasing winding number $n$ the strength of 
repulsion increases.
Plotting the above numbers over $n$ gives smooth curves. This makes us confident 
that the results obtained are not numerical errors. 
Of course, the hypothesis of repelling BI monopoles strongly relies on the agreement found
in the Abelian Higgs model and in the Georgi-Glashow model for the $n=1$ monopoles.
A direct numerical analysis of the equations is definitely called for.

\section{Conclusions}
In this paper, we have studied an approximation for
the Born-Infeld term both in the abelian as well as in the non-abelian
Higgs model. In the abelian Higgs model, we find that inserting
the data of the standard (i.e. undeformed) Nielsen-Olesen vortex in the
$\beta^{-2}$-correction we obtain a very good
approximation for the mass of BI vortices. For the $n=1$ monopoles
arising in the Georgi-Glashow model
this is also true for both the {\it ordinary} trace as well as the {\it
symmetrized} trace version. Using these results, we compute the
difference between  the {\it mass per winding number} of the $n=1$ and the $n=2$
monopole, respectively. Our results suggest that the Born-Infeld 
interaction leads to repulsion between like charged monopoles. 

\newpage
\section{Appendix}
\begin{center}
{\bf $\Delta_{bi}^{Str}$ for axially symmetric monopoles}
\end{center}
We expand the field strength tensor 
with respect to the Pauli matrices 
$\tau^{n}_{\lambda}$, $\lambda=x$, $\theta$, $\varphi$
(see (\ref{axg}))~:
\begin{eqnarray*}
F_{\mu \nu} & = & F^{(\lambda)}_{\mu \nu} \ \frac{\tau^{n}_{\lambda}}{2}
\end{eqnarray*}
The non-vanishing coefficients 
$F_{\mu\nu}^{(\lambda)}$ read \cite{hkk2}~:
\begin{eqnarray}
F_{x\theta }^{(\varphi )}
&  = & 
 - \frac{1}{x}\left(\partial_\theta H_1 + x \partial_x H_2 \right) 
\ , \nonumber \\
F_{x\varphi }^{(x)} 
& = & 
-n \frac{ \sin \theta}{x}\left( x \partial_x H_3 - H_1  H_4  \right) 
\ , \nonumber \\
F_{x\varphi }^{(\theta )} 
& = & 
n \frac{ \sin \theta}{x}\left( x \partial_x H_4 + H_1 H_3   
+ {\rm cot} \theta H_1 \right) 
\ , \nonumber \\
F_{\theta\varphi }^{(x)} 
& = & 
-n \sin \theta  \left( \partial_\theta H_3 - 1 +  H_2 H_4  
+ {\rm cot} \theta H_3  \right) 
\ , \nonumber \\
F_{\theta\varphi }^{(\theta )} 
& = &  
n \sin \theta \left( \partial_\theta H_4 - H_2 H_3   
 - {\rm cot} \theta \left( H_2 - H_4 \right)  \right) 
\ , \nonumber \\
\end{eqnarray}
and $F^{(\lambda)}_{\mu \nu} = -F^{(\lambda)}_{\nu \mu}$.\
Thus the first correction of the {\it symmetrized} trace version of 
the theory is given by~:
\begin{equation}
\Delta_{bi}^{Str}=\int \int \sin\theta d\theta dx x^2 Int^{Str}
\end{equation}
with the integrand $Int^{Str}$~:
\begin{equation}
Int^{Str}=\frac{1}{8} Str(F_{\mu\nu}F^{\mu\nu})^2=
\frac{1}{2} Str(F^2_{x\theta}g^{xx}g^{\theta\theta}+
F^2_{x\varphi}g^{xx}g^{\varphi\varphi}+F^2_{\theta\varphi}g^{\theta\theta}
g^{\varphi\varphi})^2
\end{equation}
Inserting the expressions for the field strength tensor gives:
\begin{eqnarray}
Int^{Str}&=&\frac{1}{2} Str[\frac{1}{4 x^2}
(F_{x\theta }^{(\varphi )})^2 (\tau^{n}_{\varphi})^2
+\frac{1}{4 \sin^2\theta x^2}
\{(F_{x\varphi }^{(x)})^2 (\tau^{n}_{x})^2+
(F_{x\varphi}^{(\theta)})^2 (\tau^{n}_{\theta})^2 \nonumber \\
&+& F_{x\varphi }^{(x)} F_{x\varphi}^{(\theta)}(\tau^{n}_{x}
\tau^{n}_{\theta}+\tau^{n}_{\theta}
\tau^{n}_{x})\} + \frac{1}{4 \sin^2\theta x^4}
\{(F_{\theta\varphi }^{(x)})^2 (\tau^{n}_{x})^2+
(F_{\theta\varphi}^{(\theta)})^2 (\tau^{n}_{\theta})^2 \nonumber \\
&+& F_{\theta\varphi }^{(x)} F_{\theta\varphi}^{(\theta)}( \tau^{n}_{x}
\tau^{n}_{\theta}+\tau^{n}_{\theta}
\tau^{n}_{x})\}]^2
\label{intaxbi}
\end{eqnarray}
Now, we use
\begin{equation}
Str(\tau^{n}_{\lambda_1}...\tau^{n}_{\lambda_p})=\frac{1}{p!} tr(
\tau^{n}_{\lambda_1}...\tau^{n}_{\lambda_p}+ {\rm all \ \ permutations})
\end{equation}
With this, (\ref{intaxbi}) can be evaluated in a straighforward way
by noting that
\begin{equation}
Str\left((\tau^{n}_{\lambda})^4\right)=2  \ , \ 
Str \left( (\tau^{n}_{\lambda_1})^2 (\tau^{n}_{\lambda_2})^2 \right)=\frac{2}{3}
\ , \  Str \left( (\tau^{n}_{\lambda_1})^3 (\tau^{n}_{\lambda_2}) \right)=0 \ , \
Str \left( (\tau^{n}_{\lambda_1})^2 (\tau^{n}_{\lambda_2}) 
(\tau^{n}_{\lambda_3})\right)=0
\end{equation}
where it is understood that $\lambda_1\neq \lambda_2\neq \lambda_3$.
We obtain:
\begin{eqnarray}
Int^{Str}&=&\frac{1}{16x^4} [(F_{x\theta }^{(\varphi )})^4 
+\frac{1}{ \sin^4\theta}((F_{x\varphi }^{(x)})^2+
\frac{(F_{\theta\varphi}^{(x)})^2}{x^2})^2  
+\frac{1}{ \sin^4\theta}((F_{x\varphi }^{(\theta)})^2+
\frac{(F_{\theta\varphi}^{(\theta)})^2}{x^2})^2
\nonumber \\
&+& \frac{4}{3\sin^4\theta}(F_{x\varphi }^{(x)} F_{x\varphi}^{(\theta)}
+\frac{F_{\theta\varphi }^{(x)} F_{\theta\varphi}^{(\theta)} }{x^2})^2
+\frac{2}{3\sin^2\theta}(F_{x\theta }^{(\varphi)})^2\{(F_{x\varphi }^{(x)})^2
+ \frac{(F_{\theta\varphi }^{(x)})^2 }{x^2}+
(F_{x\varphi }^{(\theta)})^2
+ \frac{(F_{\theta\varphi }^{(\theta)})^2 }{x^2}\} \nonumber \\
&+& \frac{2}{3\sin^4\theta}\{(F_{x\varphi }^{(x)})^2+
\frac{(F_{\theta\varphi }^{(x)})^2 }{x^2}\}\{(F_{x\varphi }^{(\theta)})^2+
\frac{(F_{\theta\varphi }^{(\theta)})^2 }{x^2}\}]
\end{eqnarray}

\end{document}